\documentclass[11pt]{article}

\usepackage{graphicx}
\usepackage{amsmath}
\usepackage{amsfonts}
\usepackage{amssymb}
\usepackage{verbatim}
\usepackage{hhline}

\renewcommand{\arraystretch}{1.3}

\catcode`\@=11
\def\marginnote#1{}

\newcount\hour
\newcount\minute
\newtoks\amorpm
\hour=\time\divide\hour by60 \minute=\time{\multiply\hour by60
\global\advance\minute by-\hour}
\edef\standardtime{{\ifnum\hour<12 \global\amorpm={am}%
        \else\global\amorpm={pm}\advance\hour by-12 \fi
        \ifnum\hour=0 \hour=12 \fi
        \number\hour:\ifnum\minute<10 0\fi\number\minute\the\amorpm}}
\edef\militarytime{\number\hour:\ifnum\minute<10 0\fi\number\minute}

\def\draftlabel#1{{\@bsphack\if@filesw {\let\thepage\relax
      \xdef\@gtempa{\write\@auxout{\string
          \newlabel{#1}{{\@currentlabel}{\thepage}}}}}\@gtempa \if@nobreak
    \ifvmode\nobreak\fi\fi\fi\@esphack} \gdef\@eqnlabel{#1}}
    \def\@eqnlabel{}
\def\@vacuum{}
\def\draftmarginnote#1{\marginpar{\raggedright\scriptsize\tt#1}}

\def\draft{
  \oddsidemargin -.5truein
  \def\@oddfoot{\footnotesize \sl preliminary draft \hfil
    \rm\thepage\hfil\sl\today\quad\militarytime}
  \let\@evenfoot\@oddfoot \overfullrule 3pt
    \let\label=\draftlabel
    \let\marginnote=\draftmarginnote
  \def\@eqnnum{(\theequation)\rlap{\kern\marginparsep\tt\@eqnlabel}%
    \global\let\@eqnlabel\@vacuum}

  }

\makeatletter
\newdimen\normalarrayskip              
\newdimen\minarrayskip                 
\normalarrayskip\baselineskip \minarrayskip\jot
\newif\ifold             \oldtrue            \def\new{\oldfalse}
\def\arraymode{\ifold\relax\else\displaystyle\fi} 
\def\eqnumphantom{\phantom{(\theequation)}}     
\def\@arrayskip{\ifold\baselineskip\z@\lineskip\z@
     \else
     \baselineskip\minarrayskip\lineskip2\minarrayskip\fi}
\def\@arrayclassz{\ifcase \@lastchclass \@acolampacol \or
\@ampacol \or \or \or \@addamp \or
   \@acolampacol \or \@firstampfalse \@acol \fi
\edef\@preamble{\@preamble
  \ifcase \@chnum
     \hfil$\relax\arraymode\@sharp$\hfil
     \or $\relax\arraymode\@sharp$\hfil
     \or \hfil$\relax\arraymode\@sharp$\fi}}
\def\@array[#1]#2{\setbox\@arstrutbox=\hbox{\vrule
     height\arraystretch \ht\strutbox
     depth\arraystretch \dp\strutbox
     width\z@}\@mkpream{#2}\edef\@preamble{\halign
\noexpand\@halignto
\bgroup \tabskip\z@ \@arstrut \@preamble \tabskip\z@ \cr}%
\let\@startpbox\@@startpbox \let\@endpbox\@@endpbox
  \if #1t\vtop \else \if#1b\vbox \else \vcenter \fi\fi
  \bgroup \let\par\relax
  \let\@sharp##\let\protect\relax
  \@arrayskip\@preamble}
\def\eqnarray{\stepcounter{equation}%
              \let\@currentlabel=\theequation
              \global\@eqnswtrue
              \global\@eqcnt\z@
              \tabskip\@centering
              \let\\=\@eqncr

 \halign to \displaywidth\bgroup
    \eqnumphantom\@eqnsel\hskip\@centering
    $\displaystyle \tabskip\z@ {##}$%
    \global\@eqcnt\@ne \hskip 2\arraycolsep
         $\displaystyle\arraymode{##}$\hfil
    \global\@eqcnt\tw@ \hskip 2\arraycolsep
         $\displaystyle\tabskip\z@{##}$\hfil
         \tabskip\@centering
    &{##}\tabskip\z@\cr}
\newfont{\hr}{msbm10}
\newfont{\ams}{msam10}

\hoffset= - 0.9in         

\def\beq{\begin{equation}}
\def\eeq{\end{equation}}
\def\ba{\beq\new\begin{array}{c}}
\def\ea{\end{array}\eeq}
\def\be{\ba}
\def\ee{\ea}

\def\N2{${\cal N}=2$}

\def\1N{${\cal N}=1$}
\def\4N{${\cal N}=4$}
\def\nn{\nonumber}

\def\p{\partial}

\def\la{\left\left<}
\def\ra{\right\right>}

\newdimen\linethick  \linethick=0.4pt
\newdimen\hboxitspace    \hboxitspace=5pt
\newdimen\vboxitspace    \vboxitspace=5pt

\def\fr#1{%
\beq\new \vcenter{ \hrule height\linethick
          \hbox{\vrule width\linethick
                \kern\hboxitspace
                \vbox{\kern\vboxitspace
                      \hbox{$\begin{array}{c}\displaystyle#1
         \end{array}$}%
                      \kern\vboxitspace}%
                \kern\hboxitspace
                \vrule width\linethick}%
          \hrule height\linethick}%
\eeq}

\renewcommand{\tt}[1][mer]{\hbox{\tiny{#1}}}

\def\vect{\hat}
\def\p{\partial}
\def\tr{{\rm tr}\,}

\def\p{\partial}
\def\tr{{\rm tr}\,}

\def\la{\left<}
\def\ra{\right>}
\def\lala{\la\!\!\!\!\la}
\def\rara{\ra\!\!\!\!\ra}

\hoffset= - 1.0in         

\title{{\bf
Towards a proof of AGT conjecture
by methods of matrix models} 
\vspace{.2cm}}
\author{{\bf A.Mironov}\footnote{ {\small {\it
Lebedev Physics Institute} and {\it ITEP, Moscow, Russia}};
mironov@itep.ru; mironov@lpi.ru}, {\bf A.Morozov}\thanks{{\small
{\it ITEP, Moscow, Russia} and {\it Laboratoire de Mathematiques et
Physique Theorique, CNRS-UMR 6083, Universite Francois Rabelais de
Tours, France}}; morozov@itep.ru} \ and {\bf
Sh.Shakirov}\thanks{{\small {\it ITEP, Moscow, Russia} and {\it
MIPT, Dolgoprudny, Russia}}; shakirov@itep.ru}\date{ }}

\begin{document}

\setcounter{footnote}{3}

\setcounter{tocdepth}{3}

\maketitle

\vspace{-6.cm}

\begin{center}
\hfill FIAN/TD-10/10\\
\hfill ITEP/TH-44/10
\end{center}

\vspace{5.cm}

\begin{abstract}
A matrix model approach to proof of the AGT relation is briefly
reviewed. It starts from the substitution of conformal blocks by the
Dotsenko-Fateev $\beta$-ensemble averages and Nekrasov functions
by a double deformation of the exponentiated Seiberg-Witten
prepotential in $\beta\neq 1$ and $g_s\neq 0$ directions. Establishing
the equality
of these two quantities is a typical matrix model problem, and
it presumably can be ascertained by investigation of integrability
properties and developing an associated Harer-Zagier technique for
evaluation of the exact resolvent.
\end{abstract}

\section{Introduction}

The AGT relation \cite{AGT}
experimentally discovered
about 1.5 years ago, brings together a number of different subjects
some of them listed in Fig.1. In this short review we are going to
explain them very briefly. The details and
various discussions of the AGT conjecture can be found in
\cite{AGTfirst}-\cite{MMMsurop}.
The correspondence shown in the figure,
can be described at different levels:
as representation theory,
as in the first line;
as a direct equality of two quantities
in the second line:
conformal blocks
of two-dimensional conformal theory
and Nekrasov functions of Seiberg-Witten (SW)
theory;
and as a relation between matrix models and
Seiberg-Witten theory in the third line.
The main problem with the first two options
is the lack of a reasonable
{\it conceptual} definition of
the Nekrasov functions, only operational
technical definitions are available so far.
At the third level, the Nekrasov functions
are considered as providing a double-deformation
with the help of the two $\epsilon$-parameters
of the Seiberg-Witten theory.
We find this point of view at the AGT conjecture
the most interesting and important.
It involves, at the physical level,
the correspondence between
gauge theories and integrable systems,
which are well known to stand behind
both the matrix models \cite{GMMMO}-\cite{UFN3}
and the SW theory \cite{intSW1,intSWd,intSW}.
In this paper we rely upon the standard
dictionary (gauge theory $\leftrightarrow$
integrable system) \cite{intSWd}.
Also we use the term ``matrix models''
in a wider sense:
the eigenvalue $\beta$-ensembles.
All the definitions from
$2d$ conformal theory used below
can be found in \cite{CFT}
and \cite{MMMagt}.

In a few first sections we remind
the definitions of the quantities that
are directly used in the AGT correspondence:
the conformal blocks and Nekrasov functions, and formulate the AGT conjecture.
Then we explain what definition of SW theory and of matrix models we use.
After this, we are already able to formulate the AGT correspondence
as a matrix models/SW theory correspondence in a more
concrete form (figures 5 and 6). The rest of the paper is devoted to the discussion
of its details.

At the end of the paper we add three Appendices.
Two of them are aimed to illustrate the
non-widely-known chapters of matrix model theory,
which are heavily exploited in the main text.
The first of them is the Harer-Zagier explicit formula
for the exact (all genus) one-point resolvent $\rho_1$,
which is {\it not} directly deducible from the Virasoro
constraints and, thus, from the AMM/EO version of
topological recursion; rather it is a direct corollary of a hidden
integrable structure of the free energy.
The second one is the SW representation
of the $\beta$-ensemble free energy,
$\delta F = \beta\sum_I \oint_{B_I}\rho_1
\delta \oint_{A_I}\rho_1$
through contour integrals of the exact one-point
resolvent.
We illustrate both subjects with the simplest
example of the Gaussian phase of the Hermitian matrix model
with $\beta=1$.
The program outlined in the present paper,
requires an extension of this example to
the Dotsenko-Fateev (DF) $\beta$-ensemble.
The third Appendix contains a brief description
of various limits of the AGT relation,
including the highly non-trivial
"pure gauge" and "stationary" ($\epsilon_2 \rightarrow 0$)
limits.


\bigskip

\begin{center}
\begin{figure}
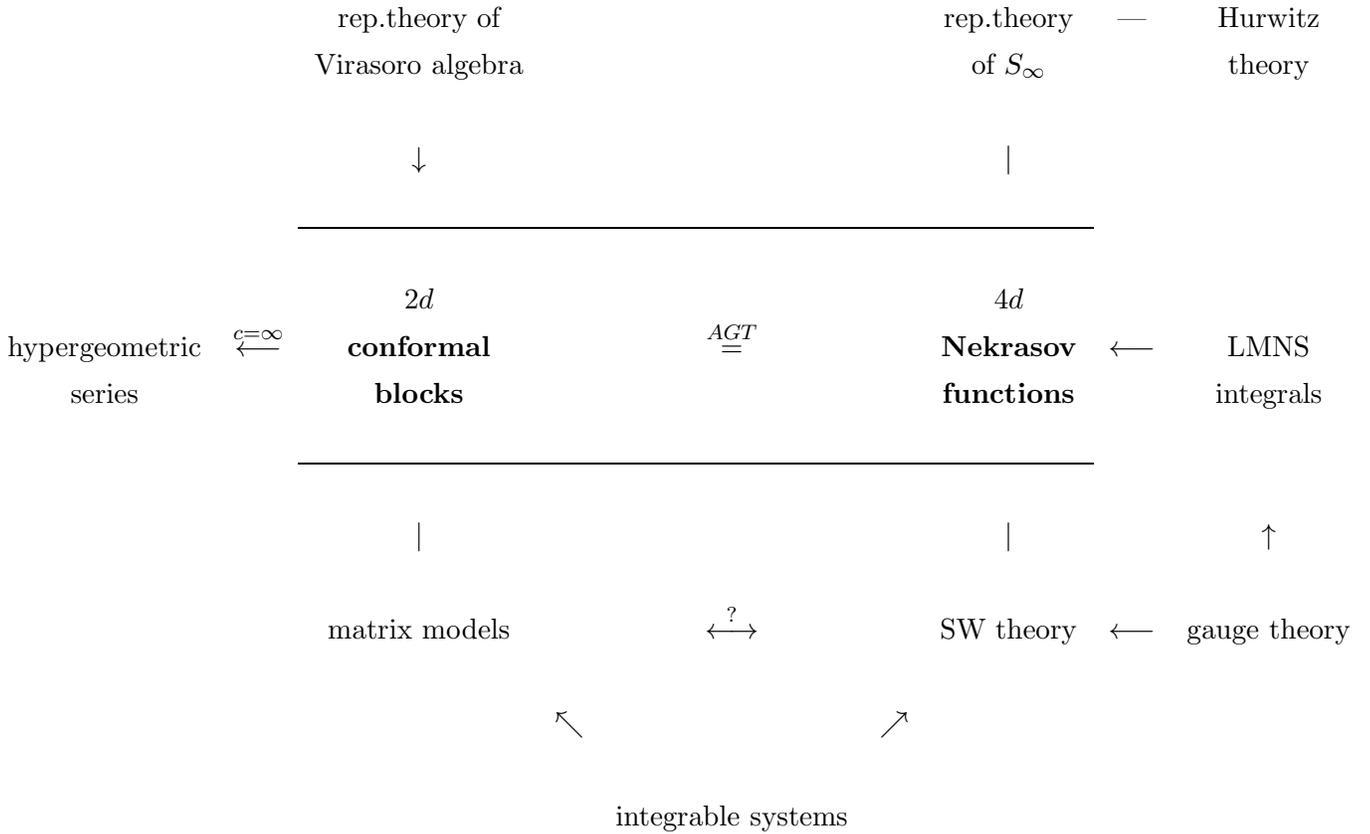

\begin{tabular}{ccccccccccc}
&& rep.theory of &&&& rep.theory & --- & Hurwitz&&\\
&&Virasoro algebra&&&& of $S_\infty$ &&theory&&\\
&&&&&&&&&&\\
&&$\downarrow$&&&& $|$ &&&&\\
&&&&&&&&&&\\
\hhline{~~|-----|~~~~}
&&&&&&&&&&\\
&&$2d$&&&&$4d$&&&&\\
hypergeometric & $\stackrel{c=\infty}{\longleftarrow}$ & {\bf conformal} &&
$\stackrel{AGT}{=}$ && {\bf Nekrasov} &
$\longleftarrow$ & LMNS && \\
series &&{\bf blocks}  &&&& {\bf functions} && integrals &&\\
&&&&&&&&&&\\
\hhline{~~|-----|~~~~}
&&&&&&&&&&\\ 
&&$|$&&&& $|$  &&$\uparrow $ \\ 
&&&&&&&&& & \\
&& matrix models &&$\stackrel{?}{\longleftrightarrow}$
&& SW theory &$\longleftarrow$& gauge theory  \\
&&&&&&&&\\
&&&$\nwarrow $&&$\nearrow $&&&\\
&&&&&&&&\\
&&&&integrable systems&&&&
\end{tabular}
\caption{Interrelations between various fields touched by the AGT
conjecture.}
\end{figure}
\end{center}

\unitlength 1mm 
\linethickness{0.4pt}
\ifx\plotpoint\undefined\newsavebox{\plotpoint}\fi 
\begin{picture}(47.94,-06.98)(-110,56)
\put(18.061,18.879){\line(1,0){22.818}}
\multiput(8.993,24.899)(.049970214,-.033593421){177}{\line(1,0){.049970214}}
\multiput(17.838,18.953)(-.058757277,-.03364733){148}{\line(-1,0){.058757277}}
\multiput(47.94,24.899)(-.039031239,-.033633302){179}{\line(-1,0){.039031239}}
\multiput(40.953,18.879)(.047206701,-.03364733){148}{\line(1,0){.047206701}}
\put(9.96,27){$V_{\Delta_2}(q)$}
\put(9.96,10){$V_{\Delta_1}(0)$}
\put(47.048,27){$V_{\Delta_3}(1)$}
\put(47.048,10){$V_{\Delta_4}(\infty)$}
\put(28.392,20.662){$\Delta$}
\end{picture}

\section{Conformal blocks}


\hangindent=-4.5cm
The 4-point conformal block is a function of the double ratio of
4 points on the Riemann sphere, $q={(z_1-z_2)(z_3-z_4)\over (z_1-z_3)(z_2-z_4)}$
which is parameterized by four external and one intermediate
conformal dimensions $\Delta_i$ ($i=1,2,3,4$), $\Delta$
and the central charge $c$ of conformal theory \cite{CFT}.
The conformal block is completely fixed by the conformal properties
and, at small enough $q$, it can be expanded into the power series:
\be
{\cal B}_{\Delta}(\Delta_1,\Delta_2;\Delta_3,\Delta_4|q) =
\sum_{n} q^n {\cal B}_{\Delta}^{(n)}
\ee
where the first coefficients of expansion are
\be
{\cal B}_\Delta ^{(0)}=1
\nn
\ee
\be
{\cal B}_\Delta^{(1)} =
{(\Delta+\Delta_1 -\Delta_2   )(\Delta+\Delta_3-\Delta_4)
\over 2\Delta}
\label{B1}
\nn
\ee
\be
{\cal B}_\Delta^{(2)}=
{(\Delta+\Delta_1 -\Delta_2   )(\Delta+\Delta_1 -\Delta_2   +1)
(\Delta+\Delta_3-\Delta_4)
(\Delta+\Delta_3-\Delta_4+1)\over 4\Delta(2\Delta+1)}+
\nn\\ \nn \\
+{\left[(\Delta_2   +\Delta_1 )(2\Delta+1)+\Delta(\Delta-1)
-3(\Delta_2   -\Delta_1 )^2\right]
\left[(\Delta_3+\Delta_4)(2\Delta+1)+\Delta(\Delta-1)
-3(\Delta_3-\Delta_4)^2\right]
\over 2(2\Delta+1)\Big(2\Delta(8\Delta-5) + (2\Delta+1)c\Big)}
\label{B2}
\nn
\ee

\bigskip

\be
\ldots
\nn
\ee

\bigskip

These expressions become fast quite involved. However, at the particular
case of large central charge $c\longrightarrow \infty$ the conformal block drastically
simplifies becoming the hypergeometric function
\be
{\cal B}_{\Delta}\Big (\Delta_1,\Delta_2;\Delta_3,\Delta_4|q\Big )\  \stackrel{ c \longrightarrow \infty}{\longrightarrow}\
F\Big( \Delta+\Delta_1-\Delta_2,\ \Delta + \Delta_3-\Delta_4;\ 2\Delta \Big | q\Big)
\ee

\section{The origin of conformal blocks}

If one writes the conformal block as a correlator in a chiral algebra with
a projector to the intermediate state inserted
\be
{\cal B}_{\Delta}(\Delta_1,\Delta_2;\Delta_3,\Delta_4|q)
= \ \Big< V_{\Delta_1(0)} V_{\Delta_2}(q) P_\Delta
V_{\Delta_3}(1) V_{\Delta_4}(\infty) \Big>
\ee
its generic structure can be determined using the operator product expansion
of two primary chiral vertices
\be
V_{\vect\alpha}(z)V_{\vect\beta}(z') =
\sum_{\vect\gamma}
\frac{{{\cal C}_{\vect\alpha\vect\beta}^{\vect\gamma}}
\,V_{\vect\gamma}(z')}
{(z-z')^{\Delta_{\vect\alpha}+\Delta_{\vect\beta}
-\Delta_{\vect\gamma}}}
\label{OPE}
\ee
of the Virasoro algebra
\be
[L_m,L_n] = (m-n)L_{m+n} + \frac{c}{12}n(n^2-1)\delta_{n+m,0}
\ee
The conformal dimension of the vertex is given by
\be
L_0 V_\alpha = \Delta_\alpha V_\alpha
\ee
and the descendant operators are defined by
\be
V_{\alpha,Y}
= L_{-k_l}\ldots L_{-k_2}L_{-k_1} V_\alpha
\label{desc}
\ee
Then, the conformal block is manifestly given by the formula
\be
{\cal B}_{\Delta}(\Delta_1,\Delta_2;\Delta_3,\Delta_4|q) =
\sum_{|Y|=|Y'|} q^{|Y|} {\cal B}_{Y,Y'}
= \sum_{|Y|=|Y'|} q^{|Y|}\beta_{\Delta_1\Delta_2}^\Delta(Y)
Q_\Delta(Y,Y')\beta_{\Delta_3\Delta_4}^\Delta(Y')
\\
=\sum_{|Y|=|Y'|} q^{|Y|}\gamma_{\Delta_1\Delta_2\Delta}(Y)
Q^{-1}_\Delta(Y,Y')
\gamma_{\Delta_3\Delta_4\Delta}(Y')
\label{calBdef}
\ee
where the 3-vertices are manifestly given by formulas
\be
\gamma_{\Delta_1,\Delta_2;\Delta}(Y) =
\langle L_{-Y}V_\Delta(0) V_{\Delta_1}(1)V_{\Delta_2}(\infty)\ \rangle
= \prod_i \left( \Delta + k_i\Delta_1 - \Delta_2
+  \sum_{j< i} k_j\right)
\label{gammaprod}
\ee
and the Shapovalov matrix is defined as a scalar product
\be
Q_\Delta(Y,Y') = \langle L_{-Y}V_{\Delta}(0)\ L_{-Y'}V_\Delta(\infty)\rangle
\label{Q2p}
\ee

\section{Nekrasov functions\label{NF}}

The Nekrasov functions for ${\cal N}=2$ SUSY $SU(2)$  gauge theory with four
fundamental matter hypermultiplets (the $\beta$-function in such a theory is equal
to zero) at first levels are manifestly given by formulas (see \cite{NF} for
generic expressions)
\be
{\cal Z}_{[1][0]} = -\frac{1}{\epsilon_1\epsilon_2}\cdot
\frac{\prod_{r=1}^4 (a + \mu_r)}
{2a(2a+\epsilon)},\nn\\
{\cal Z}_{[0][1]} = -\frac{1}{\epsilon_1\epsilon_2}\cdot
\frac{\prod_{r=1}^4 (a - \mu_r)}
{2a(2a-\epsilon)};
\label{Z1}
\ee
\be
{\cal Z}_{[2][0]} = \frac{1}{2!\,\epsilon_1\epsilon_2^2
(\epsilon_1-\epsilon_2)}\cdot
\frac{\prod_{r=1}^4 (a + \mu_r)(a+\mu_r+\epsilon_2)}
{2a(2a+\epsilon_2)(2a+\epsilon)(2a+\epsilon+\epsilon_2)},\nn\\
{\cal Z}_{[0][2]} = \frac{1}{2!\,\epsilon_1\epsilon_2^2
(\epsilon_1-\epsilon_2)}\cdot
\frac{\prod_{r=1}^4 (a - \mu_r)(a-\mu_r-\epsilon_2)}
{2a(2a-\epsilon_2)(2a-\epsilon)(2a-\epsilon-\epsilon_2)},\nn\\
{\cal Z}_{[11][0]} = -\frac{1}{2!\,\epsilon_1^2\epsilon_2
(\epsilon_1-\epsilon_2)}\cdot
\frac{\prod_{r=1}^4 (a + \mu_r)(a+\mu_r+\epsilon_1)}
{2a(2a+\epsilon_1)(2a+\epsilon)(2a+\epsilon+\epsilon_1)}, \nn \\
{\cal Z}_{[0][11]} = -\frac{1}{2!\,\epsilon_1^2\epsilon_2
(\epsilon_1-\epsilon_2)}\cdot
\frac{\prod_{r=1}^4 (a - \mu_r)(a-\mu_r-\epsilon_1)}
{2a(2a-\epsilon_1)(2a-\epsilon)(2a-\epsilon-\epsilon_1)}, \nn\\
{\cal Z}_{[1][1]} = \frac{1}{\epsilon_1^2\epsilon_2^2}\cdot
\frac{\prod_{r=1}^4 (a + \mu_r)(a-\mu_r)}
{(4a^2-\epsilon_1^2)(4a^2-\epsilon_2^2)};
\label{Z2}
\ee
\be
{\cal Z}_{[3][0]} = -\frac{1}{3!\,\epsilon_1\epsilon_2^3
(\epsilon_1-\epsilon_2)(\epsilon_1-2\epsilon_2)}\cdot
\frac{\prod_{r=1}^4 (a + \mu_r)(a+\mu_r+\epsilon_2)(a+\mu_r+2\epsilon_2)}
{2a(2a+\epsilon_2)(2a+2\epsilon_2)(2a+\epsilon)(2a+\epsilon+\epsilon_2)
(2a+\epsilon+2\epsilon_2)},\nn\\
\ldots
\label{Z3}
\ee
Here $$\epsilon = \epsilon_1+\epsilon_2$$

\section{An origin of Nekrasov functions}

The origin of the Nekrasov functions is the
LMNS integrals \cite{LMNS}, which are
integrals over a regularized moduli space of instantons, that is,
the space of ADHM configurations. These integrals are manifestly given by
formulas
\be
Z = e^F = \sum_k
q^k \oint \prod_{a<b}^k \frac{\chi_{ab}^2\Big(\chi_{ab}^2-(\epsilon_1-\epsilon_2)^2\Big)}
{\Big(\chi_{ab}^2-\epsilon_1^2\Big)\Big(\chi_{ab}^2-\epsilon_2^2\Big)}
\prod_{a=1}^k \frac{P(\chi_a)}{Q(\chi_a)}\, d\chi_a
= \sum_{\hbox{\scriptsize over pairs of}\atop\hbox{\scriptsize Young diagrams}} Z_{Nek}(Y,Y')
\ee
with $\chi_{ab} = \chi_a-\chi_b$ and the polynomials
\be
P(x)=\prod_{a=1}^4(x+\mu_a)\ ,\ \ \ \ \ Q(x)=(x^2-a^2)((x+\epsilon)^2-a^2)
\ee
depend on the vacuum expectation value $a$ of the scalar in the gauge theory
and fundamental hypermultiplet masses $\mu_a$.
These integrals can be manifestly taken and
represented as sums over Young diagrams (integer partitions) \cite{NF}.

The associated brane diagrams are the tropical limits of $2d$ Riemann surfaces.
The correspondence between the conformal blocks and the brane
diagrams is drawn on figures 2 and 3.

\begin{figure}
\begin{picture}(100,30)(-30,-10)
\put(0,0){\line(1,0){100}}
\put(20,0){\line(0,1){20}}
\put(40,0){\line(0,1){20}}
\put(60,0){\line(0,1){20}}
\put(80,0){\line(0,1){20}}
\put(2,-5){\makebox(0,0)[cc]{$0$}}
\put(15,18){\makebox(0,0)[cc]{$x$}}
\put(85,18){\makebox(0,0)[cc]{$1$}}
\put(98,-5){\makebox(0,0)[cc]{$\infty$}}
\end{picture}
\caption{The diagram describing the multi-point conformal block of the ``comb''
type, which is involved into the AGT relation.}
\end{figure}
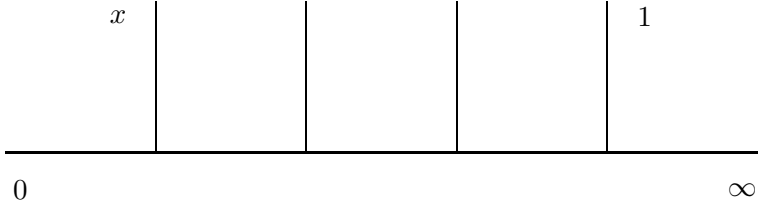

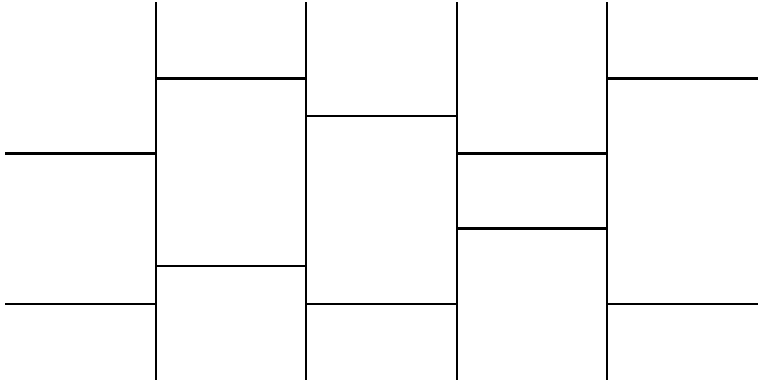
\begin{figure}
\begin{picture}(100,60)(-30,-125)
%
\put(150,5){\makebox(0,0)[cc]{$S-duality$}}
%
%
%
\put(20,-120){\line(0,1){50}}
\put(40,-120){\line(0,1){50}}
\put(60,-120){\line(0,1){50}}
\put(80,-120){\line(0,1){50}}
\put(0,-110){\line(1,0){20}}
\put(0,-90){\line(1,0){20}}
\put(20,-105){\line(1,0){20}}
\put(20,-80){\line(1,0){20}}
\put(40,-110){\line(1,0){20}}
\put(40,-85){\line(1,0){20}}
\put(60,-100){\line(1,0){20}}
\put(60,-90){\line(1,0){20}}
\put(80,-110){\line(1,0){20}}
\put(80,-80){\line(1,0){20}}
\end{picture}
\caption{The brane diagram AGT related to the comb conformal block in Fig.2.}
\end{figure}

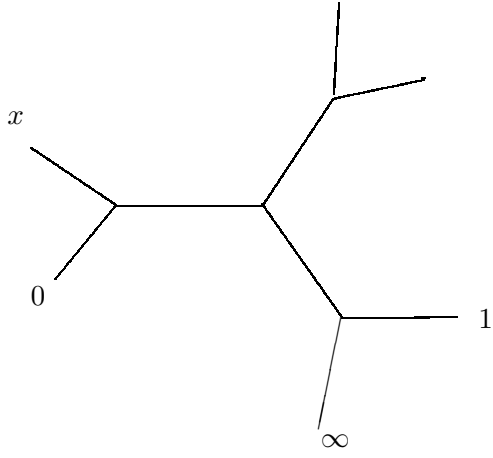
\begin{figure}
\begin{picture}(50,60)(50,70)
\multiput(129.25,109)(.0336363636,.0509090909){275}{\line(0,1){.0509090909}}
\multiput(138.5,123)(.16,.03333333){75}{\line(1,0){.16}}
\multiput(150.5,125.5)(-.03125,.03125){8}{\line(0,1){.03125}}
\put(150.25,125.75){\line(1,0){.5}}
\put(150.75,125.75){\line(0,1){0}}
\multiput(139.25,135.75)(-.0326087,-.5326087){23}{\line(0,-1){.5326087}}
\multiput(129,109)(.0336990596,-.0478056426){319}{\line(0,-1){.0478056426}}
\multiput(139.75,93.75)(1.90625,.03125){8}{\line(1,0){1.90625}}
\put(139.5,94){\line(-1,-5){3}}
\put(129,108.75){\line(-1,0){19.5}}
\multiput(109.5,108.75)(-.033613445,-.040966387){238}{\line(0,-1){.040966387}}
\multiput(109.75,108.75)(-.05,.033695652){230}{\line(-1,0){.05}}
\put(99.25,96.75){\makebox(0,0)[cc]{0}}
\put(96.25,120.5){\makebox(0,0)[cc]{$x$}}
\put(158.75,93.75){\makebox(0,0)[cc]{1}}
\put(138.75,77.5){\makebox(0,0)[cc]{$\infty$}}
\end{picture}
\caption{One of the diagrams obtained from the comb diagram of Fig.2 by
an S-duality transformation.\label{new}}
\end{figure}

Here we pictured the correspondence for the comb type diagrams.
The other type of diagrams are
obtained from this one by a transformation of S-duality, see Fig.\ref{new}.
The elements of matrix of
the same
transformation at the level of representation theory (of Virasoro algebra)
are named the Racah coefficients (while the conformal block is nothing but
the Clebsch-Gordan coefficients, which depend on the parameter $q$, since in
the Virasoro case the co-product is parameterized by this parameter \cite{MS}).

\section{AGT relation}

The AGT conjecture literally states 
that
\be
{\cal B}_{\Delta}(\Delta_1,\Delta_2;\Delta_3,\Delta_4|q) =
\sum_{n} x^n {\cal B}_{\Delta}^{(n)} =
\sum_{|Y|=|Y'|} q^{|Y|}
{\cal B}_{Y,Y'}^\alpha(\alpha_1,\alpha_2;\alpha_3,\alpha_4) =
\nn
\ee
\be
=
(1-q)^{-\nu} \sum_{Y,Y'} q^{|Y|+|Y'|} {\cal Z}^{SU(2)}_{Y,Y'}
\equiv \sum_{Y,Y'} q^{|Y|+|Y'|} {\cal Z}^{U(2)}_{Y,Y'}
\label{BvsZ}
\ee
where ${\cal Z}^{SU(2)}_{Y,Y'}$ are given by the formulas of section \ref{NF}.

The simplest example of the conjecture comes from the first level (linear term in $q$):
\be
{\cal B}^{(1)}_\Delta
= \frac{(\Delta + \Delta_1-\Delta_2)(\Delta + \Delta_3-\Delta_4)}{2\Delta} =
\nn \\ \nn \\
= {\cal Z}_{[1][0]} + {\cal Z}_{[0][1]} + \nu =
-\frac{1}{\epsilon_1\epsilon_2}\cdot
\frac{\prod_{r=1}^4 (a + \mu_r)}{2a(2a+e)}
-\frac{1}{\epsilon_1\epsilon_2}\cdot
\frac{\prod_{r=1}^4 (a - \mu_r)}{2a(2a-e)} + \nu
\label{BZ1}
\ee
which is correct provided that
\be
\Delta_i = \frac{\alpha_i(\epsilon  - \alpha_i)}{\epsilon_1\epsilon_2}
\label{dime}
\ee
\be
a+\frac{\epsilon}{2} = \alpha
\ee
\be\label{solutions}
\mu_1=-{\epsilon\over 2}+\alpha_1+\alpha_2, \ \ \mu_2={\epsilon\over
2}+\alpha_1-\alpha_2, \ \ \mu_3=-{\epsilon\over
2}+\alpha_3+\alpha_4, \ \ \
\mu_4={\epsilon\over 2}+\alpha_3-\alpha_4,
\ee
\be
\nu = \frac{2\alpha_1\alpha_3}{\epsilon_1\epsilon_2}
\ee
At higher levels one also needs
\be
c = 
1 + \frac{6(\epsilon_1+\epsilon_2)^2}{\epsilon_1\epsilon_2} = 1 -
6\left(b-\frac{1}{b}\right)^2
\label{ce}
\ee
with
\be\label{betags}
b = \sqrt{-\epsilon_1/\epsilon_2}, \ \ \ \ \
g_s = \sqrt{-\epsilon_1\epsilon_2}
\ee

\section{Seiberg-Witten representation of prepotential}

We call SW prepotential a function ${\cal F}(a_1,\ldots,a_n)$ of $n$ variables
defined with the help of the SW data:
a complex curve $\Sigma$ and an
meromorphic 1-differential
$\Omega$ on it, through the SW system of equations:

\be
\left\{\begin{array}{c}
a_I = \oint_{A_I} \Omega \\\\
\frac{\partial {\cal F}}{\partial a_I} = \oint_{B_I} \Omega
\end{array}\right.
\label{SWeq}
\ee

\noindent
In our study it appears in two contexts.

\paragraph{I. SW theory associated with the $SU(2)$ gauge model.}
The prepotential ${\cal F}_0^{SU(2)}(a)$ of a single variable $a$
is defined on $\Sigma^{SU(2)}$ of genus one and \cite[(1.1)]{Arg}
\be\label{SWXXX}
\Omega_0^{SU(2)} = pd\phi,
\\
\Sigma^{SU(2)}:\ \ \ \ \ \ \ \ {p^2} - P_+(p)e^{i\phi}-P_-(p)e^{-i\phi}  = E
\ee
where
\be
P_+(p-h)={(p-\mu_{1})(p-\mu_{2})\over (1+q)},\ \ \ \ \ \
P_-(p-h)={q(p-\mu_3)(p-\mu_4)\over (q+1)},\ \ \ \ \ \
h\equiv {\sum_{i=1}^4\mu_i\over 2(q+1)}
\ee
and $q$
is a function of the coupling constant $\tau$, $q=\theta_4^4/\theta_2^4$
(see \cite{Arg,MMMzam}).
This SW system corresponds to the
XXX spin chain on 2 sites \cite{intSWd}.
The limit of all four hypermultiplet masses infinite (see s.\ref{puregauge})
leads to the pure gauge theory on the physical side \cite{SW1},
to the periodic Toda chain
on 2 sites on the integrable side \cite{intSWd} and to a degenerated conformal block
on the conformal side \cite{nonconf}. In this case,
the corresponding SW pair (\ref{SWXXX}) is
\be\label{SWToda}
\Omega^{SU(2),pg}_0 = pd\phi\\
\Sigma^{SU(2),pg}:\ \ \ \ \ \ \ \ \frac{1}{2}p^2 - \Lambda^2\cos \phi = E
\ee
The prepotential ${\cal F}_0^{SU(2)}(a)$
is lifted up to the Nekrasov function ${\cal F}^{SU(2)}(a)$ so that
${\cal F}^{SU(2)}(a)\to {\cal F}_0^{SU(2)}(a)$ in the limit
$\epsilon_1=-\epsilon_2\to 0$.

\paragraph{II. SW theory associated with the matrix model.}
In any eigenvalue matrix model one can define a resolvent
and its genus expansion in powers of $g_s$ (or $1/N$) \cite{AMM1}-\cite{EO}
(see s.13 for
accurate definitions, the sign minus in front of the SW differential is due to the
sign minus in front of the matrix model potential, (\ref{betai})):
\be
-\Omega^{(mm)} =
\rho^{(mm)}(z) = \left<{\rm Tr}\frac{1}{z-M}\right>_{(mm)}
= \sum_{p=0}^\infty g_s^{2p}\rho_p^{(mm)}(z)
\ee
All $\rho_p^{(mm)}(z)$ are meromorphic 1-differentials on
a spectral curve $\Sigma^{(mm)}$ and the free energy
${\cal F}^{(mm)} = \sum_{p=0}^\infty g_s^{2p}{\cal F}_p^{(mm)}$
satisfies the SW equations, (\ref{SWeq}). We illustrate how formulas
(\ref{SWeq}) work in the simplest case of the Gaussian matrix model in
Appendix II.

In fact, along with the SW pair $\Big({\cal F}^{(mm)}, \ \rho^{(mm)}\Big)$,
one also can associate with each matrix model
the "quasiclassical" (genus zero) SW pair
$\Big({\cal F}_0^{(mm)}, \ \rho_0^{(mm)}\Big)$.

As any partition function,
the exponential of ${\cal F}^{(mm)}$ is a $\tau$-function,
satisfying {\it bilinear} Hirota equations \cite{Hirota},
while ${\cal F}_0^{(mm)}$ is a far more sophisticated
"quasiclassical" $\tau$-function, satisfying a system of highly
non-linear WDVV equations \cite{WDVV}
$${F}_I{F}_J^{-1}{F}_K = {F}_K{F}_J^{-1}{F}_I$$
where $n\times n$ matrix $\left({F}_I\right)_{JK}
\equiv \frac{\partial^3{\cal F}}{\partial a_I\partial a_J\partial a_k}$ .

\section{DF and Selberg matrix models}

The generic Virasoro conformal block can be considered
as an analytic continuation in $N_1$ and $N_2$ of
the DF type integrals \cite{MMS1,MMS2}:
\be
{\cal B}_{\Delta}(\Delta_1,\Delta_2;\Delta_3,\Delta_4|q) = \nn
\ee
\be
\left<
:e^{\alpha_1\phi(0)}:\
:e^{\alpha_2\phi(q)}:\
:e^{\alpha_3\phi(1)}:\
:e^{\alpha_4\phi(\infty)}:\
\left(\int_0^q
:e^{b\phi(x)}:\,dx\right)^{N_1}
\left(\int_0^1
:e^{b\phi(y)}:\,dy\right)^{N_2}
\right>_{{\rm free\ fields}} \sim \nn
\ee
\be\label{beta}
\sim
q^{2\alpha_1\alpha_2}(1-q)^{2\alpha_2\alpha_3}
\prod_{i=1}^{N_1}\int_0^q dx_i \prod_{k=1}^{N_2}\int_0^1 dy_k
\prod_{i<j}^{N_1}(x_i-x_j)^{2b^2}
\prod_{k<l}^{N_2}(y_k-y_l)^{2b^2}
\prod_{i,k}^{N_1}(x_i-y_k)^{2b^2}\cdot\\ \cdot
\prod_i x_i^{2\alpha_1b}(q-x_i)^{2\alpha_2b}(1-x_i)^{2\alpha_3b}
\prod_k y_k^{2\alpha_1b}(q-y_k)^{2\alpha_2b}(1-y_k)^{2\alpha_3b}
\sim \\ \sim
\left<\left< \exp \left\{2\sum_{m=1}^\infty \frac{q^m}{m}
\left(\alpha_2+b\sum_i^{N_1} x_i\right)
\left(\alpha_3 + b\sum_k^{N_2} y_k\right)\right\}
\right>_{N_1}\right>_{N_2}
\ee

\bigskip

\noindent
Integrals of this kind are often called the $\beta$-ensemble, $\beta=b^2$
being the degree of the Van-der-Monde determinant in the integrand.

The l.h.s. is a free chiral field correlator,
evaluated with the help of the Wick theorem,
and the r.h.s. is a double average in two Selberg models,

\be
\Big<f(x)\Big>_{N_1} \equiv\
S_{N_1}^{-1}\int_0^1 \ f(x)\prod_{i<j}^{N_1}(x_i-x_j)^{2b^2}
\prod_{i=1}^{N_1}x_i^{2\alpha_1b}(1-x_i)^{2\alpha_2b}dx_i  ,\nn \\
\Big<g(y)\Big>_{N_2} \equiv\ S_{N_2}^{-1}\int_0^1
\ g(y)
\prod_{k<l}^{N_2}(y_k-y_l)^{2b^2}
\prod_{k=1}^{N_2}
y_k^{2\alpha_1b}(1-y_k)^{2\alpha_3b} dy_k
\ee

\bigskip

These Selberg integrals are explicitly evaluated and are rational
functions of $\alpha$ and $b$ parameters,
decomposable into linear factors
whenever $f$ and $g$ are the Jack polynomials $J^{(b^2)}$,
see \cite{Itomms,MMMsel} for details and references.

\section{A detailed scheme}

Now we are ready to draw a more detailed scheme as compared with Fig.1.
With the notions introduced and defined above, the new picture looks like
Fig.5. Below we briefly discuss the meaning of the three links with question marks
in the low part of the table.
If clarified,
together with the already established Dotsenko-Fateev (DF)
representation of conformal blocks
these arrows would provide
one possible proof of the AGT conjecture, the one
based essentially on the matrix model technique.
The structure of such a proof is outlined on another figure, Fig.6.

\begin{figure}
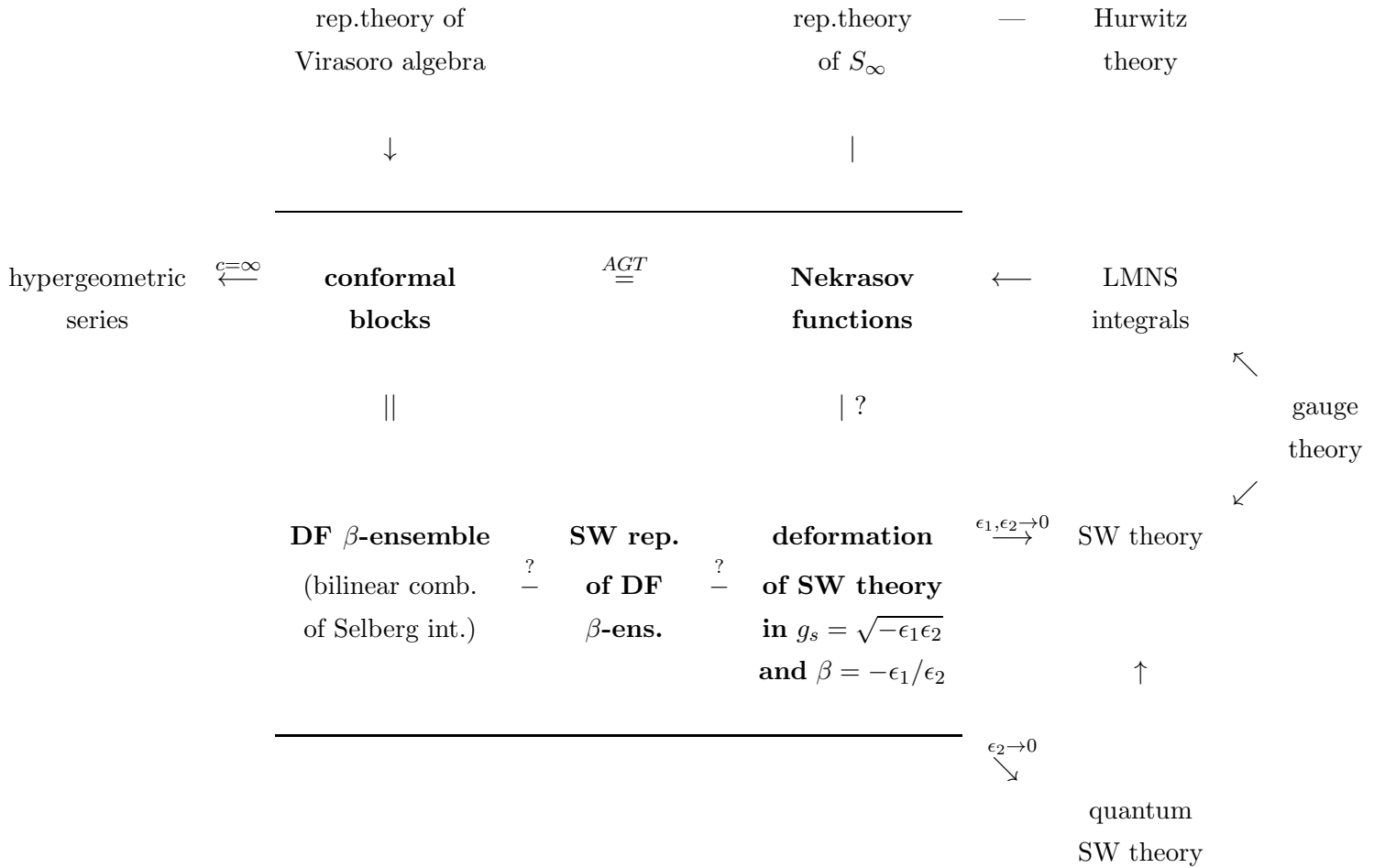

\vspace{-1cm}
\hspace{-1cm}\begin{tabular}{ccccccccccc}
&& rep.theory of &&&& rep.theory & --- & Hurwitz&&\\
&&Virasoro algebra&&&& of $S_\infty$ &&theory&&\\
&&&&&&&&&&\\
&&$\downarrow$&&&& $|$ &&&&\\
&&&&&&&&&&\\
\hhline{~~|-----|~~~~}
&&&&&&&&&&\\
hypergeometric & $\stackrel{c=\infty}{\longleftarrow}$ & {\bf conformal} &&
$\stackrel{AGT}{=}$ && {\bf Nekrasov} &
$\longleftarrow$ & LMNS && \\
series &&{\bf blocks} &&&& {\bf functions} && integrals &&\\
&&&&&&&&&$\nwarrow$&\\
&&$||$&&&& $|$ ? &&&&gauge\\
&&&&&& &&&&theory\\
&&&&&&&&&$\swarrow$&\\
&& {\bf DF $\beta$-ensemble} && {\bf SW rep.} && {\bf deformation} &
$\stackrel{\epsilon_1,\epsilon_2\rightarrow 0}{\longrightarrow}$&
SW theory&&\\
&&(bilinear comb. & $\stackrel{?}{-}$& {\bf of DF} &$\stackrel{?}{-}$&
{\bf of SW theory} &&&&\\
&&of Selberg int.)&&{\bf $\beta$-ens.}&& {\bf in} $g_s=\sqrt{-\epsilon_1\epsilon_2}$&&&&\\
&&&&&& {\bf and} $\beta=-\epsilon_1/\epsilon_2$&&$\uparrow$&&\\
&&&&&&&&&&\\
\hhline{~~|-----|~~~~}
&&&&&&&
$\stackrel{\epsilon_2\rightarrow 0}{\!\!\!\searrow}$&&&\\
&&&&&&&&quantum&&\\
&&&&&&&&SW theory&&
\end{tabular}
\caption{The detailed diagram of the AGT conjecture.}
\end{figure}

\begin{figure}
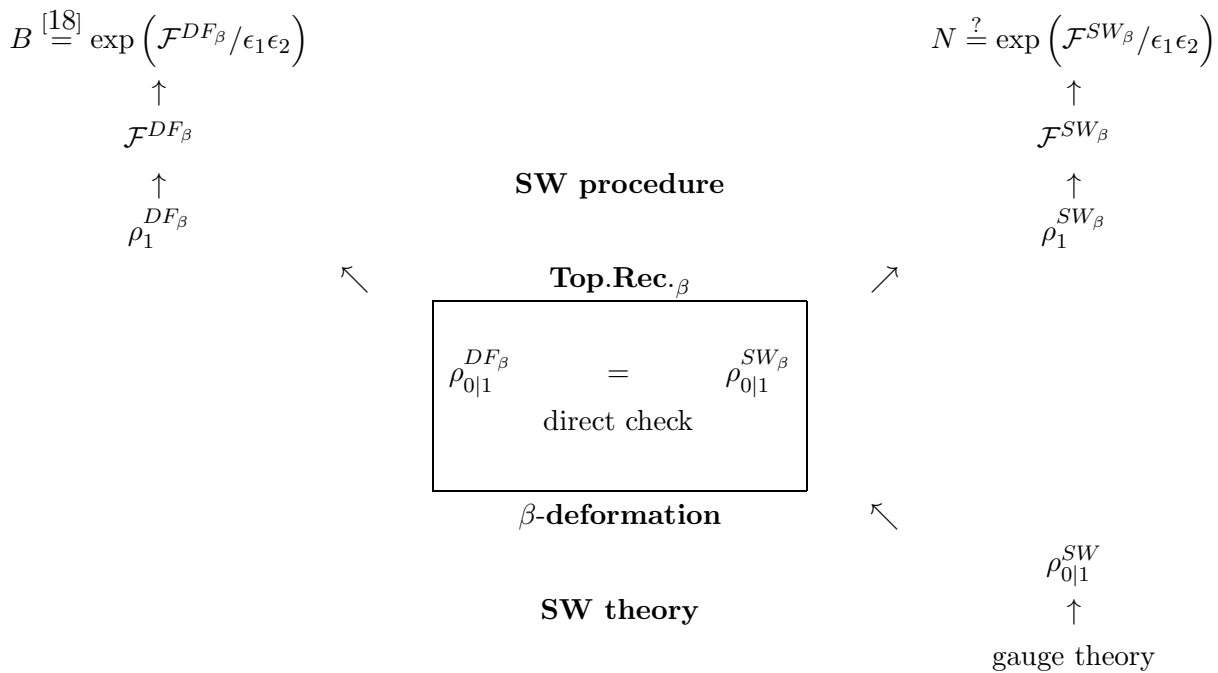

\begin{tabular}{ccccccccc}
$B\stackrel{\cite{MMS2}}{=}\exp\Big({\cal F}^{DF_\beta}/\epsilon_1\epsilon_2\Big)$&&&&&&
$N\stackrel{?}{=}\exp\Big({\cal F}^{SW_\beta}/\epsilon_1\epsilon_2\Big)$ &&\\
$\uparrow$ &&&  &&& $\uparrow$ &&\\
${\cal F}^{DF_\beta}$&&&&&& ${\cal F}^{SW_\beta}$ &&\\
$\uparrow$ &&& {\bf SW procedure} &&& $\uparrow$ &&\\
$\rho_1^{DF_\beta}$ &&&&&& $\rho_1^{SW_\beta}$ && \\
&$\nwarrow$ && ${\bf Top.Rec.}_\beta$ && $\nearrow$ &&&\\
&&&\begin{tabular}{|ccc|}
\hline
&&\\
$\rho_{0|1}^{DF_\beta}$ &=& $\rho_{0|1}^{SW_\beta}$\\
&{\rm direct\ check}&\\
&&\\
\hline
\end{tabular} &&&&\\
&&&$\beta$-{\bf deformation}&&$\nwarrow$&&&\\
&&&&&&$\rho_{0|1}^{SW}$\\
&&&{\bf SW theory}&&&$\uparrow$&\\
&&&&&&gauge theory\\
\end{tabular}
\caption{The scheme of the proof of the AGT conjecture.}
\end{figure}

The proof of the AGT relation between the conformal block $B$
and the Nekrasov function $N$ is here actually substituted
with a proof of a much simpler identity:
between the genus zero resolvents associated with the
$\beta$-deformed
spectral curves of the (Dotsenko-Fateev)matrix and Seiberg-Witten models,
\be
\rho_{0|1}^{DF_\beta} = \rho_{0|1}^{SW_\beta}
\label{01rel}
\ee
This relation (\ref{01rel}) can be established by a direct check,
which for $\beta=1$ has been performed in \cite{Eguchi1}.
For $\beta\neq 1$ one still needs to know the $\beta$-deformed version
of $\rho_{0|1}^{SW_\beta}$, which is provided by the generic
procedure of $\beta$-deformation of the spectral curves,
see s.\ref{betadef} below.
Once (\ref{01rel}) is established, one lifts it by two {\it canonical}
operations first to the equality of the {\it full} resolvents,
\be
\rho_{1}^{DF_\beta} = \rho_{1}^{SW_\beta}
\ee
with the help of the ($\beta$-deformed) topological recursion
and, second, to equality of the free energies
\be
{\cal F}^{DF_\beta} = {\cal F}^{SW_\beta}
\ee
with the help of the standard Seiberg-Witten procedure,
which builds up the prepotential ${\cal F}$ from the Seiberg-Witten
differential $\rho_1$.
Finally, exponentiating the free energy provides the conformal
blocks and the Nekrasov functions.
In the case of conformal blocks, this equality,
\be
B\stackrel{\cite{MMS2}}{=}\exp\Big({\cal F}^{DF_\beta}/\epsilon_1\epsilon_2\Big)
\ee
is nothing but a
{\it new} Dotsenko-Fateev type representation of conformal blocks
with {\it open} integration contours
introduced in \cite{MMS2}.
In the case of Nekrasov functions this statement,
\be
N\stackrel{?}{=}\exp\Big({\cal F}^{SW_\beta}/\epsilon_1\epsilon_2\Big)
\ee
remains to be proved, and the possibility of giving such a proof
depends very much on the choice of a proper {\it definition} of Nekrasov functions:
for various definitions see \cite{LMNS,NFv}.

In the remaining part of this paper we briefly comment on
the main ingredients of this suggested proof, especially on the three
{\it canonical} operations: $\beta$-deformation, topological recursion
(i.e. $g_s$-deformation) and the SW procedure.
All formulas below are given for the simplest
4-point spherical conformal block.
Generalizations to arbitrary conformal blocks,
at least, spherical and topic,
and associated quiver gauge theories are straightforward.

\section{Topological recursion}

Switching on the string coupling $g_s\neq 0$, i.e. the deformation
of the quasiclassical
$${\cal F}_0^{(mm)} = {\cal F}^{(mm)}(g_s=0)$$
into the full ${\cal F}^{MAMO},$
has actually a {\it functorial} description,
with no reference to matrix models.
This lifting
$$(\Sigma,\Omega_0) \longrightarrow \Omega = \sum_{p=0}^\infty g_s^{2p}\Omega_p$$
or, equivalently,
$$(\Sigma,\Omega_0) \longrightarrow
\tau{\rm-function},$$
or even a "quantization" $\tau_{quasicl}\rightarrow \tau$,
is now known under the name of {\it topological recursion},
where "topological" refers to the "genus expansion" in powers of $g_s$.
It already has a numerous applications to different subjects,
not explicitly related to matrix models.

\bigskip

The well-publicized part of the story, the AMM/EO construction
\cite{AMM1}-\cite{EO}
consists of building up a hierarchy of poly-differentials
(multi-resolvents) on the spectral curve $\Sigma$.
It is a part of a more general theory, \cite{AMM.IM},
covering also {\it the decomposition formulas},
i.e. the construction of integrability preserving intertwiners
(of which the simplest example are $W$-representations of \cite{MSw,Alexw}).
This approach "implicitly" refers to the Virasoro constraints, i.e.
the matrix model Ward identities \cite{Vir}, and is very general.
However, it is not very practical if one needs to construct $\Omega$,
because it involves an auxiliary construction of all unneeded,
and far more complicated, {\it multi}-resolvents at
intermediate steps.

\bigskip

An alternative approach \cite{HZ,AMM1,MS1,MS2,Haagerup}
builds
$$\Omega = \sum_{p=0}^\infty g_s^{2p}\Omega_p$$
directly from $\Omega_0$, by writing and solving the difference-differential
equation for $\Omega$, which involves recursion over $N\sim 1/g_s$
(actually, over the matrix size), and follows from integrability properties of
$\tau = e^{\cal F}/g_s^2$.
It is explicitly known only in the case of Hermitean model,
but there is nothing preventing
one from developing a similar technique of
the Harer-Zagier (HZ) recursion,
for other matrix models.

\paragraph{Topological recursion: global description.}
Now we describe the topological recursion a bit more concretely.
The construction implies that
the whole set of multi-resolvents is
intimately related to the $\widehat{U(1)}$ current $\hat{\cal J}(z)$
on the spectral curve $\Sigma$, with prescribed singularities: usually they are
allowed at some fixed points (\textit{punctures}) on $\Sigma$.
In this approach the Virasoro constraints on partition function
are written as
\be\label{proj}
\hat{\cal P}_-\left(\hat{\cal J}^2(z)\right) Z \equiv
\hat{\cal J}* \hat{\cal J}Z=
\oint_C {\cal K}(z,z') \left(\hat{\cal J}^2(z)\right) Z = 0
\ee
with a certain kernel ${\cal K}(z,z')$, made out of the
free-field Green function on $\Sigma$.
The current is also "shifted":
$\hat{\cal J}(z) \longrightarrow \hat{\cal J}(z) + \Delta\hat{\cal J}(z)$
and partition function $Z$ depends on the choice of:

$\bullet$ the complex curve (Riemann surface) $\Sigma$,

$\bullet$ the Green function ${\cal K}(z,z')$, i.e. projection operator
$\hat{\cal P}_-$,

$\bullet$ the punctures on $\Sigma$ and associated loop operator
$\hat{\cal J}(z)$,

$\bullet$ the local coordinates in the vicinity of the punctures,

$\bullet$ the involution of the curve with punctures and loop operator,

$\bullet$ the shift $\Delta{\cal J}(z)$ on $\Sigma$,

$\bullet$ the contour $C$ which separates two sets of punctures.

If contour $C$ goes around an isolated puncture, $Z$ is actually
defined by its infinitesimal vicinity and depends on behavior
(the type of singularity) of $\hat {\cal J}(z)$ at this particular
puncture. Coordinate dependence is reduced to the action of
a unitary operator (Bogoliubov transform, and exponential of bilinear
function of $\hat{\cal J}$) on $Z$.

If contour $C$ is moved away from the vicinity of the puncture,
it can be decomposed into contours encircling all other punctures:
this provides relations between $Z$'s of different types,
associated with different punctures, these are exactly the decomposition formulas.

\paragraph{Topological recursion: local description.}
The standard recursive loop equations \cite{AMM1}-\cite{EO}
are reproduced from this global construction locally in the vicinity of a fixed point.
To this end, one has to choose the local parameter $z$ in this vicinity and put
\be
\hat{\cal J}=V' + g_s^2\hat\nabla
\ee
where
\be
\hat\nabla(z) = \sum_{k=0}^\infty
\zeta_k(z) \frac{\partial}{\partial T_k} \label{nabla}
\ \ \ \ \ \ \
V'(z)
= \sum_{k=0}^\infty \tilde T_k v_k(z)
\ee
and $v_k(z)$ and $\zeta_k(z)$ are the full sets of 1-forms on $\Sigma$,
related by the condition
\be
\hat\nabla(z) V'(z') = B(z,z')
\label{Berg}
\ee
where $B(z,z')$ is the Bergmann kernel, i.e.
$(1,1)$ Green function $B(z,z') = <\partial\phi(z)
\partial\phi(z')>$ on $\Sigma$. The integral in (\ref{proj})
gives rise to a multiplication map $\Omega_\Sigma\times\Omega_\Sigma
\rightarrow \Omega_\Sigma$ on the space of 1-forms $\Omega_\Sigma$.
For hyperelliptic curves, which are double coverings of the Riemann
sphere, the contour $C$ is a finite set of contours encircling
the ramification points and $\tilde
z$ is the counterpart of $z$ on the other sheet. Then the kernel $K$
is actually a differential of the form $\frac{dz}{d\tilde z'}$,
which is a ratio of the $(1,0)$ Green function on $\Sigma$ (which is
the primitive of the Bergmann kernel w.r.t. the second argument
calculated from $z'$ to $\tilde z'$) and the
Seiberg-Witten-Dijkraaf-Vafa differential \footnote{In simplest
case of the sphere, $\Sigma_H:\ y^2_H(z)=z^2-4S$ corresponding to the Hermitean
one-matrix model \cite{AMM.IM}
$$
K(z,z')= \frac{dz}{dz'}\frac{1}{z-z'}\left(\frac{1}{y_H(z)} -
\frac{1}{y_H(z')}\right)
$$}:
\be K(z,z') = \frac{<\partial\phi(z)\, \phi(z')>-<\partial\phi(z)\,
\phi(\tilde z')>}{\Omega_{DV}(z')-\Omega_{SW}(\tilde z')} \ee See
\cite{AMM.IM} and \cite{CMMV,EO} for details.

Substitution of (\ref{nabla}) into (\ref{proj})
gives:
\be
\left( g_s^2\sum_{k,n\geq 0}(v_k*\zeta_n) \tilde
T_k\frac{\partial}{\partial T_n} + \frac{g_s^4}{2}\sum_{k,l\geq
0}(\zeta_k*\zeta_l) \frac{\partial^2}{\partial T_k\partial T_l} +
\frac{1}{2}\sum_{k,l\geq 0} (v_k*v_l) \tilde T_k\tilde T_l +{1\over
2}Tr_*B \right) Z=0 \label{AMME2}
\ee
where the shift of the current $\Delta{\cal J}(z)$
is in charge of the $*$-trace of the
Bergmann kernel. Expanding the products of 1-forms into linear
combinations of $\zeta$ (no $v$ will arise due to projection
property of the $*$-product), one obtains a one-dimensional set of
constraints on $\log Z\equiv\sum_{p\ge 0} g_s^{2p-2}{\cal F}^{(p)}$.
They can be also written as recurrent relations
for the multiresolvents
\be
\rho^{(p|m)}(z_1,\ldots,z_m) = \left.
\hat\nabla(z_1)\ldots\hat\nabla(z_m){\cal F}^{(p)}\right|_{T_k=\delta_{k,1}}
\ee
in the following form
\be \rho^{(p|m+1)}(z,z_1,\ldots,z_m) =
\frac{1}{2}Tr_* B(\bullet,\bullet) \delta_{p,0}\delta_{m,0} +
\sum_{i=1}^m B(\bullet,z_i)*\rho^{(p|m)}(\bullet,z_{I/i}) + \nn \\ +
\sum_{p_1=0}^p\sum_{J\subset I}
\rho^{(p_1|m_J+1)}(\bullet,z_J)*\rho^{(p_2|m_{I/J}+1)}(\bullet,I/J)
+ \frac{1}{2}Tr_* \rho^{(p-1|m+2)}(\bullet,\bullet,z_1,\ldots,z_m)
\ee
They are obtained simply by acting with operators $\hat\nabla$
on (\ref{proj}) and putting $T_k-\delta_{k,1}=0$
afterwards. The terms with the Bergmann kernel come from the action
of $\hat\nabla$ on $V'$, action on the $V'*V'$ term gives rise to
the trace of the Bergmann kernel. The notation here is as follows:
the bullets, $\bullet$ mark arguments on which the
$*$-product acts, two points are converted into a single $z$. If
both bullets are arguments in the same function, we call the
corresponding product $*$-trace, $Tr_*$: for, say, $H(z_1,z_2) =
\sum_{m,n} H_{mn}\zeta_m(z_1)\zeta_n(z_2)$ the $*$-trace is
$Tr_*H(\bullet,\bullet) = \sum_{m,n} H_{mn}(\zeta_m* \zeta_n)(z)$.

\section{$\beta$-deformation\label{betadef}}

$\beta$-ensembles differ from eigenvalue matrix models \cite{UFN3}
by substitution of the second power of the Van-der-Monde determinant
$\Delta(M)$ in the measure on the space of eigenvalues by
arbitrary power $2\beta$.
This leads to a straightforward, but rather sophisticated
deformation of integrals involving only traces and
determinants of the integration $\beta$-ensemble variables.
The $\beta$-character calculus \cite{Unint,PGL}
which can be used to describe this deformation,
automatically provides also the tools to handle the
eigenvalue models with one external
field, like Kontsevich model \cite{Ko}, GKM \cite{GKM}
and BGWM \cite{BGW}.
Character calculus in \cite{PGL} is not, however,
developed enough to describe $\beta$-deformations of the
Kontsevich and DV phases in these models as well as
of their integrability properties.
In the context of {\it this} paper
it is sufficient to know the $\beta$-deformations
of the spectral curve and Virasoro constraints
and thus of the AMM/EO topological recursion.
Remarkably, these are the simplest things to
$\beta$-deform, and one does need to know anything
about the $\beta$-character calculus.
In particular, formula (\ref{beta}) describes the
$\beta$-ensemble with $\beta=b^2$.
This is actually a simple deformation:
it affects the spectral curve $\Sigma$ only through rescaling
of coefficients, and it slightly changes the AMM/EO topological
recursion (in particular, for $\beta\neq 1$ the $g_s$-expansion
also includes {\it odd} powers of $g_s$, (\ref{WI1}): contributions from
open and non-oriented surfaces, "halves of their closed surface
doubles").

\paragraph{$\beta$-deformation of the SW data.}
The SW data (\ref{SWeq}) for the $\beta$-ensemble is trivially deformed
\be
\left\{\begin{array}{c}
a_I = -\oint_{A_I} \rho_1 \\\\
\frac{\partial {\cal F}}{\partial a_I} =-\beta \oint_{B_I} \rho_1
\end{array}\right.
\label{SWeqb}
\ee
The genus zero
one-point resolvent for $\beta$-ensembles looks like (see s.13 for the details)
\be
-\beta\rho_{0|1}^2 + W'\rho_{0|1} = f(z) \equiv \Big(W'\rho\Big)_+
\ \ \ \ \ \ \ \
\rho_{0|1}(z) =  \frac{W'(z) - \sqrt{W'(z)^2 - 4\beta f(z)}}{2\beta}dz
\ee
and the spectral curve is
\be\label{38}
\left(\rho_{0|1}(z)-{W'(z)\over 2\beta}\right)^2=y^2(z)=
{W'(z)^2 - 4\beta f(z)\over 4\beta^2}
\ee
For the generic SW theory
one should first represent $\rho_{0|1}$ at $\beta=1$
in this form with some $W'(z)$ and $f(z)$ and then introduce $\beta$.
For example, for the $SU(2)$ pure gauge theory (\ref{SWToda})
(here we subtracted the potential part from the genus zero resolvent, see
s.13)
\be
\rho_{0|1}^{SU(2)} = {\sqrt{\Lambda^2 \cos z - E}\over\sqrt{2}}\,dz =
\sqrt{\left(\Lambda \cos\frac{z}{2}\right)^2 - E'}\,dz
\ \ \ \longrightarrow \ \ \
\rho_{0|1}^{SU(2)_\beta} =
\frac{1}{\beta}\sqrt{\left(\Lambda \cos\frac{z}{2}\right)^2 - \beta E'}
\ee

\paragraph{$\beta$-deformation of the Virasoro algebra (=loop equations).}
The AMM/EO topological recursion is easily $\beta$-deformed.
For example, the Virasoro operators
for Hermitean model change from
\be
-\frac{\partial}{\partial t_{n+2}} +
\sum_k kt_k\frac{\partial}{\partial t_{k+n}}
+ \sum_{a+b=n} \frac{\partial^2}{\partial t_a\partial t_b}
\ee
to
\be
-\frac{\partial}{\partial t_{n+2}} +
\sum_k kt_k\frac{\partial}{\partial t_{k+n}}
+ \beta \sum_{a+b=n} \frac{\partial^2}{\partial t_a\partial t_b}
+(1-\beta)(n+1)\frac{\partial}{\partial t_n}
\ee

More generally, one constructs the $\beta$-deformation of the topological recursion
completely along the line of the previous section, with only two ingredients deformed.
First of all, one has to deform the spectral curve
(and, certainly, the related quantities like
${\cal K}(z,z')$ and the Bergmann kernel) and, then, the Sugawara construction, which is
behind (\ref{proj}). The latter deformation is well-known and can be read off immediately
using the standard conformal matrix model technique \cite{confmamo,Ito}: one has to
replace (\ref{proj}) with
\be
\hat{\cal P}_-\left(\hat{\cal J}^2(z)+\left(b-{1\over b}\right)\p_z \hat{\cal J}
\right) Z =0
\ee

\bigskip

It is yet unknown how does the $\beta$-deformation affects the
integrability-inspired HZ topological recursion,
but one can also assume that the changes will not be too drastic.

\section{The LMNS double deformation of SW theory}

As clear from consideration of the previous sections, one
can write the {\it full} free energy of the $\beta$-ensemble as
\be
{\cal F}^{(mm),\beta} =
{\rm top.rec.}_\beta\Big[ {\cal F}_0^{(mm),\beta}\Big]
\ee
where the r.h.s. involves only {\it functorial} operation
of topological recursion and the quasiclassical (genus zero)
free energy. This means that the r.h.s. is fully defined
by nothing but the SW pair $(\Sigma,\Omega_0)$.

Moreover, the same operation can be applied
to the arbitrary SW pair $(\Sigma,\Omega^{SW}_0)$
to obtain a $\beta,g_s$ double deformation
\be
{\cal F}^{SW,\beta} =
{\rm top.rec.}_\beta\Big[ {\cal F}_0^{SW,\beta}\Big]
\ee

The Nekrasov functions have several different definitions,
see \cite{LMNS,NFv} for some of them.
One of the definitions implies that they arise as a double
deformation of the (exponentiated) gauge theory SW prepotential
by two parameters $\epsilon_1,\epsilon_2$.
This double deformation can presumably be alternatively described
as switching on $\beta\neq 1$ and $g_s\neq 0$ with the usual
AGT identifications (\ref{betags}), $\beta=b^2$.
In other words, one can define the Nekrasov functions as
\be
\log Z_{Nek}^{SU(2)} = {\cal F}^{SU(2),\beta} =
{\rm top.rec.}_\beta\Big[ {\cal F}_0^{SU(2),\beta}\Big]
\ee
where at the r.h.s. stands just the usual
gauge theory SW prepotential.

\section{Identification of SW and $\beta$-ensemble differentials
$\Omega_0$ and spectral curves}

Now in order to establish the AGT conjecture one has to check first that the
SW data coincides in the both cases: in the SW theory and in the
$\beta$-ensemble. Then, applying the topological recursion, one lifts the both
to the full free energies which then also coincide.

\paragraph{SW data in $\beta$-ensembles.} Assume one starts from the
$\beta$-ensemble
\be\label{betai}
Z\sim\int \prod_i dz_i \exp\left(-{1\over g_s}W(z_i)\right)
\prod_{i<j}(z_i-z_j)^{2\beta}
\ee
Then, using the invariance under the shift of integration variables
$\delta z_i \sim z_i^{n+1}$ with $n\geq -1$, one immediately obtains
(the double brackets here mean integration with the $\beta$-ensemble
measure (\ref{betai}))
\be
\lala 2\beta g_s\sum_{i<j}\frac{z_i^{n+1}-z_j^{n+1}}{z_i-z_j}
- \sum_i W'(z_i)z_i^{n+1}
+ g_s\sum_i nz_i^n\rara
\ = 0
\ee
Summing up these identities over $n$ with the weights $\xi^{-n-2}$,
one rewrites these Virasoro constraints in the form of a loop equation
for the one-point resolvents:
\be
\beta\rho_1^2(\xi)-W'(\xi)\rho_1(\xi)+f(\xi)=(1-\beta)g_s{\p\rho_1(\xi)\over\p\xi}
-\beta g_s^2\rho_2(\xi,\xi)
\label{WI1}
\ee
where
\be
\rho_k(\xi_1,\ldots,\xi_k)\equiv g_s^k\lala\prod_{a=1}^k\sum_{i}{1\over \xi_a
-z_i}
\rara_c\ ,\ \ \ \ \ \ \ f(\xi)\equiv
\lala\sum_i{W'(\xi)-W'(z_i)\over \xi -z_i}\rara
\ee
and the subscript $c$ means the connected correlator.
Solving the loop equation (\ref{WI1})
in the leading order in the string coupling constant $g_s$,
one obtains (compare with (\ref{38}), the sign is chosen here in order to
guarantee the correct behaviour at large $\xi$)
\be
\rho_{0|1}(\xi)={W'(\xi)-\sqrt{W'^2(\xi)-4\beta f(\xi)}\over 2\beta}
\ee
In fact, it is usually more convenient
to define $\rho_{0|1}(z)$ with the potential part subtracted, not changing
the definition of higher order resolvents,
and from now
on we use the one-point resolvent defined exactly in such a way.
The spectral curve and the SW differential are then
\be\label{51}
\Omega_0 = ydz\\
\Sigma:\ \ \ \ \ \ \ y^2={W'^2(\xi)-4\beta f(\xi)\over 4\beta^2}
\ee

\paragraph{DF integrals as $\beta$-ensembles.}
Choosing now the potential
\be
W(\xi)=-2\epsilon_1\sum_{k=1}^3\alpha_k\log(\xi-q_k),\ \ \ \ \ \
\ \ \ \ \{q_1,q_2,q_3\}=\{0,q,1\}
\ee
one reproduces the $\beta$-ensemble (\ref{beta}) and obtains (after rescaling $y$)
from (\ref{51})
the following SW data (see
\cite[(3.10)]{Eguchi1}, \cite[(3.38)]{Eguchi2}, \cite[(46)]{MMS1}):
\be\label{SWbeta4m}
\Omega_0 = ydz\\
\Sigma:\ \ \ \ \ \ \ y^2={M_-^2\over z^2}+{M_+^2\over (z-1)^2}
+{m_+^2\over (z-q)^2}-{M_+^2+M_-^2+m_+^2-m_-^2\over z(z-1)}-{(1-q)u\over z(z-1)(z-q)}
\ee
Here
\be
m_\pm=\alpha_{2,4}={\mu_1\pm\mu_2\over 2},\ \ \ \ \ \ M_\pm=\alpha_{3,1}=
{\mu_3\pm\mu_4\over 2}
\ee

Note that the same spectral curve can be obtained via absolutely different procedure
of studying the differential equation satisfied by the conformal block with
a degenerate field inserted \cite[(2.24)]{MT}, \cite[(88)]{MMMsurop}.

\paragraph{SW theory vs. planar limit of the DF $\beta$-ensemble.}
In order to compare the SW data (\ref{SWbeta4m})
with that of the corresponding SW theory, eq.(\ref{SWXXX}),
one has to make the change of variables in this latter
\be
e^{i\phi}=z,\ \ \ \ \ \ \ \left(p-{(\mu_1+\mu_2-2h)e^{2i\phi}+q(\mu_3+\mu_4-2h)\over
2(e^{i\phi}-1)(e^{i\phi}-q)}\right)e^{-i\phi}=y
\ee
which, indeed, leads to the SW pair (\ref{SWbeta4m}) upon the identification
\be
\ \ \ \ \ \
u={1+q\over 1-q}E-{(M_+-m_+)^2\over 1-q^2}+{q\mu_1\mu_2+\mu_3\mu_4\over 1-q}
\ee
Thus, we prove that the SW data for the DF $\beta$-ensemble
and the $SU(2)$ gauge theory with four fundamental matter hypermultiplets
coincide. Then they would keep to coincide after the topological recursion
applied, i.e. the conformal block is equal to the Nekrasov functions
(defined by the topological recursion).

\section{The pure gauge limit \label{puregauge}}

The most familiar is the gauge SW theory in the "pure gauge" case,
when the four fundamental masses are taken to infinity, while
$q\rightarrow 0$, so that
$\Lambda^4 = qm_1m_2m_3m_4$ remains finite.
Then the SW pair, the spectral curve and the differential become (\ref{SWToda}).
In order to see this one suffices to make the change of variables in
(\ref{SWToda})
\be
e^{i\phi}=z,\ \ \ \ \ pe^{-i\phi}=y
\ee
and immediately to come to
\be
\Omega^{SU(2),pg}_0 = ydz\\
\Sigma^{(SU(2),pg}:\ \ \ \ \ \ \ y^2 = {\Lambda^2\over z}+{E\over z^2}+
{\Lambda^2\over z^3}
\ee
This spectral curve is exactly the curve which describes the pure gauge theory
limit on the conformal side.

Note that the DF $\beta$-ensemble in the pure gauge limit looks rather
different from (\ref{beta}). This limit is obtained by presenting the $\beta$-ensemble
integral (\ref{beta}) as an average of the product of two Selberg integrals
and further taking the limit of these Selberg integrals.
According to \cite{PGL} the final answer is given by
\be
B_*(\Delta|\Lambda) \ = \
\lim_{
\stackrel{
\stackrel{\Delta_1,\Delta_2,\Delta_3,\Delta_4\rightarrow \infty, \ q\rightarrow 0}{}
}
{q(\Delta_2-\Delta_1)(\Delta_3-\Delta_4) \equiv \Lambda^4}
}
\Big[ B(\Delta_1,\Delta_2,\Delta_3,\Delta_4;\Delta,c|q) \Big] =\\
=\exp\left( -2\beta \sum\limits_{k =
1}^{\infty} \dfrac{\Lambda^{4k}}{k} \dfrac{\partial^2}{\partial
t^{+}_k \partial t^{-}_k} \right) Z_{*}\big(n_+| t^{+}\big)
Z_{*}\big(n_-| t^{-} \big)\Big|_{t=0}
\label{decoB*}
\ee
where $Z_{*}\big( n \big| t \big)$ is the partition function of the
Brezin-Gross-Witten model in the character phase,
\be
Z_{*}\Big(n \big| t_k =
\tr \Psi^k/k \Big) = \dfrac{1}{{\rm Vol}_\beta(n)} \int\limits_{n \times n}
[dU]_{\beta} e^{\tr U^{+} + \tr \Psi U} \label{IntClaim}
\ee
and
\be
n_{\pm} = \pm \dfrac{2a}{\epsilon_1}, \ \ \ a = \alpha -
\dfrac{\epsilon_1+\epsilon_2}{2}
\ee

\section{Conclusion}

In this paper we presented a possible strategy for
proving the AGT relation between conformal blocks
and Nekrasov functions, based on existence of the
{\it canonical} double deformation of arbitrary
Seiberg-Witten prepotentials with the help of the
topological recursion.
This recursion can be applied both to the gauge theory
SW prepotentials and to the Dotsenko-Fateev free energies.
Since the spectral curves and SW differentials
in these two cases are known to coincide,
this actually implies the AGT relation.

To make this argument into a constructive identification,
with all steps explicit, a better understanding
is desirable
of the HZ version of the topological recursion,
and thus of integrability properties of non-Gaussian
and $\beta$-deformed matrix models.

\section*{Acknowledgements}

Our work is partly supported by Ministry of Education and Science of
the Russian Federation under contract 02.740.11.0608, by RFBR
grants 10-02-00509-a (A.Mir.), and 10-02-00499 (A.Mor. \& Sh.Sh.),
by joint grants 09-02-90493-Ukr, 09-01-92440-CE, 09-02-91005-ANF,
10-02-92109-Yaf-a. The work of A.Morozov was also supported in part
by CNRS.

\newpage

\section*{Appendix I. On HZ topological recursion}

The celebrated Hermitian model resolvent \cite{AMM1} (to make a contact with the standard
matrix model notations, in this Appendix
we use the definition of the resolvent without subtracting the potential term and
rescale $y$ by a factor of 2)
\be
\rho_N^{HM}(x) = \frac{x-y}{2} + \frac{N}{y^5} + \frac{21N(x^2+N)}{y^{11}}
+ \frac{11N(135z^4+558Nz^2+158N^2)}{y^{17}(z)} + \ldots
\label{serrho}
\ee
with $y^2 = x^2-4N$ actually satisfies the difference equation \cite{AMM1}
\be
\rho_{N+1} - 2 \rho_N + \rho_{N-1} = \rho^{''}_N / N
\label{HZeqRecursive}
\ee
which can be also rewritten as the differential equation \cite{MS1,Haagerup}
\be
-\rho_N''' + y^2\rho_N'- x\rho_N + 2N = 0
\label{HZeq1}
\ee
Equation (\ref{HZeqRecursive}) (or, equivalently, (\ref{HZeq1})) is very different
from the usual topological recursion. Still, it completely reproduces the asymptotic
series (\ref{serrho}). As explained in \cite{AMM1,MS1}
such equations follow from integrability of the model, in this particular case from
the lowest Toda-chain equation \cite{GMMMO}
\be
\frac{Z_{N+1}^{HM}Z_{N-1}^{HM}}{Z_N^{HM}} =
\frac{\partial^2}{\partial t_1^2} \log Z_N^{HM}
\label{Toda}
\ee
where
\be
Z_N^{HM} = \dfrac{\int\limits_{n\times n} d M e^{-\tr M^2/2 + \sum_k t_k \tr M^k} }{\int\limits_{n\times n} d M e^{-\tr M^2/2}}
\ee
is the partition function of the Gaussian Hermitian matrix model. As is also
explained in \cite{MS1} (and even earlier in \cite{AMM1}) it is often very convenient
to consider not the matrix model at particular value of $N$, but the generating
function w.r.t. $N$, the "grand-ensemble" partition function
\be
Z^{HM}(\lambda) = \sum_{N=0}^\infty \lambda^N Z_N^{HM}
\ee
and its derivatives, such as the "grand-ensemble" resolvent:
\be
\hat\rho(x|\lambda) = \sum_{N=0}^\infty \lambda^N\rho(x|N)
\label{hatrho}
\ee
This function is in many respects more clever
than its finite $N$ counterpart: just like $\rho_N$ it is given by an asymptotic series
(see eq.(42) in \cite{MS2}),
but this time the series can be explicitly summed into an error function \cite{AMM1}:
\be
\hat\rho^{HM}(x|\lambda) = \sum_{k=0}^\infty\frac{\lambda(1+\lambda)^k}
{(1-\lambda)^{k+2}}\frac{(2k-1)!!}{x^{2k+1}} =
\dfrac{i\lambda}{(1-\lambda)\sqrt{1-\lambda^2}}\
\mbox{erf}\left( i z \dfrac{1 - \lambda}{1 + \lambda} \right)
\label{allgen}
\ee
The function $\hat\rho$ satisfies
\be
\lambda \partial_\lambda \left( \dfrac{(1-\lambda)^2}{\lambda} \hat\rho \right) = \hat\rho^{''}
\ee
which is obviously equivalent to (\ref{HZeqRecursive}), or
\be
-\hat\rho''' + (x^2 - 4\lambda\partial_\lambda)\hat\rho'
- x\hat\rho + \frac{2\lambda}{(1-\lambda)^2} = 0
\ee
which is obviously equivalent to (\ref{HZeq1}).
See also \cite{HZ,Haagerup,AMM1,MS2} for more details. Integrability inspired
equations of this type are often useful in various applications of matrix models
(and, perhaps, will be useful in applications to the AGT conjecture).
To illustrate their usefulness, in the next Appendix we use eq.(\ref{HZeqRecursive})
to prove the Seiberg-Witten representation of the Gaussian Hermitian model free energy
exactly (in all genera).

\section*{Appendix II. Seiberg-Witten representation
of the Gaussian-model partition function}

The SW representation (\ref{SWeq}) of the matrix model free energy
is not known widely enough, despite its extreme conceptual importance.
It is only briefly mentioned even in recent matrix model reviews,
see, for example, \cite{MMSW,CMMV,AMM}.
Moreover, its general prove is not actually available at the moment.

In this Appendix we demonstrate how this representation works in
the simplest case of the Gaussian Hermitian model (with $\beta=1$).
In this case
\be
Z_N = \dfrac{1}{V_N} \int\limits_{N \times N} dM e^{-\frac{1}{g}\tr M^2}
= \dfrac{g^{N^2/2}}{V_N} \int\limits_{N \times N} dM e^{- \tr M^2} = g^{N^2/2} \prod\limits_{k=1}^{N-1} k!
\ee
where $V_N$ is the volume of the unitary group.
Thus
\be
F_N = \log Z_N = \frac{N^2}{2}\log g + \sum_{k=1}^{N-1} \log (k!)
\ee
This free energy actually gets contributions from all
genera, since in genus expansion one uses the
variables $S = gN$ and $g$. Equivalently, one can keep
$g = 1$ and consider the genus expansion simply as an
asymptotical expansion at large $N$. Convenient for this purpose
is the Euler-McLaurin summation formula \cite{WW}
\be
\sum_{k=0}^{N-1} f(k) = \int\limits_{0}^{N} f(x) dx
+ \dfrac{B_1}{1!} \big( f(N) - f(0) \big)
+ \sum_{k=2}{B_{k}\over k!}\left[f^{(k-1)}(N)-f^{(k-1)}(0)\right]
\ee
i.e.
\be
\frac{\p}{\p N}\sum_{k=1}^{N-1} f(k) =
f(N) + \dfrac{B_1}{1!}f'(N) + \dfrac{B_2}{2!}f''(N) +\dfrac{B_4}{4!}f''''(N) + \ldots
\label{EMf}
\ee
where $B_k$ are the Bernoulli numbers \cite{GR},
$\sum_{k=0} B_k x^k/k! = x/(e^x-1)$, $B_0=1,\ B_1=-1/2,\ B_2 = 1/6,\ B_3=0,\ B_4=-1/30,\ \ldots$
Applying (\ref{EMf}) with
\be
f(z) = \Gamma(z+1) = (z+1/2)\log z - z + \log\sqrt{2\pi} +
\sum_{m=1}^\infty \frac{B_{2m}}{2m(2m-1)z^{2m-1}}
\ee
one finds
\be
\dfrac{\partial F_N}{\partial N} = \big( N \log (gN) - N \big)
+ \log\sqrt{2\pi} - \dfrac{1}{12 N} + \dfrac{1}{120N^3} - \dfrac{1}{252N^5} + \ldots=
\\=\big( N \log (gN) - N \big) + \log\sqrt{2\pi}
- \sum\limits_{g = 1}^{\infty} \dfrac{B_{2g}}{2g} \dfrac{1}{N^{2g-1}}
\label{dfdn}
\ee
Note that only the odd powers of $1/N$ survive in the series at the r.h.s.
To obtain (\ref{dfdn}) one makes use of the bilinear identity between the
Bernoulli numbers, $\sum_{i+j=k} \frac{(2k-2)!}{(2i)!(2j)!}B_{2i}B_{2j}
= - \frac{1}{2k}B_{2k}$, $k>1$.

The aim of this appendix is to reproduce this expansion from the one-point resolvent
(\ref{serrho})
\be
\rho_1(z|N) \ \stackrel{\cite{AMM1}}{=}\
-\frac{y(z)}{2} + \frac{N}{y^5(z)} + \frac{21N(z^2+N)}{y^{11}(z)}
+ \frac{11N(135z^4+558Nz^2+158N^2)}{y^{17}(z)} + \ldots
\label{rho1}
\ee
where $y^2(z) = z^4-4N$. The $A$-period of $\rho_1$ is simple:
\be
a = -\oint_A \rho_1(z|N) dz = -\oint_{-\sqrt{4N}}^{\sqrt{4N}} \rho_1(z|N) dz = N
\ee
(hereafter, we use a peculiar notation $\oint_a^b$ for the integral
along the contour which goes around the points $a$ and $b$).
Only the first (genus-zero) term in (\ref{rho1})
contributes to the $A$-period, because
the integration contour can be taken to infinity
and $\rho^{(1|p)}\sim z^{-4p-1}$ as $z\rightarrow\infty$.
The $B$-period is more complicated:
\be
\frac{\p F}{\p a} = -\oint_B \rho_1(z|N) dz =
-\oint_{\sqrt{4N}}^\infty \rho_1(z|N) dz
\ee
It can even seem that the contributions of higher
genera diverge, since, for example, the integral
$\int_{-\sqrt{4N}}^{\infty} \frac{dz}{(z^2-4N)^{k/2}}$
along the ray $(-\sqrt{4N},\infty)$
diverges when $k\geq 0$.
This is, however, not the case:
integrals of $\rho_1$ are actually finite for $k\geq 1$,
because the integration is along the contour surrounding
the points $\sqrt{4N}$ and $\infty$, not along a
segment or a ray, thus, the contour can be taken away from the
singularities.
This works exactly in the same way as in the case of
the integral
$\int_{-\sqrt{4N}}^{\sqrt{4N}} \frac{dz}{(z^2-4N)^{k/2}}$,
the same phenomenon also ensures the finiteness of
the corrected Bohr-Sommerfeld periods in \cite{MMbz}.

Actually, making the change of variables
$\frac{z}{\sqrt{4N}}=\frac{2-\zeta}{\zeta}$,
one reduces the integral to the $B$-function,
\be
\frac{1}{2}\oint_0^1 \zeta^{a-1} (1-\zeta)^{b-1}d\zeta =
\frac{\Gamma(a)\Gamma(b)}{\Gamma(a+b)}
\ee
in the following way:
\be
\oint_{\sqrt{4N}}^\infty \frac{dz}{(z^2-4N)^{p/2}} =
\frac{2^{p+1}}{ (4N)^{p/2}} \oint_0^1 \zeta^{p-2} (1-\zeta)^{-p/2} d\zeta
= \frac{2^{p+1}}{ (4N)^{p/2}} \frac{\Gamma(p-1)\Gamma(1-p/2)}{\Gamma(p/2)}
\label{integ}
\ee
To see why the integral does {\it not} diverge, say
for $a\geq 0$, one can deform the contour into
a dumbbell, when the circle around $\zeta=0$
has a small radius $r$.
Then the contribution from the segment diverges as
$\frac{r^a}{2a}(1-e^{2\pi i a})$, while the integral
along the left circle is
$\frac{r^a}{2}\oint_0^{2\pi}e^{ia\phi}d\phi =
\frac{r^a }{2ia}(e^{2\pi i a}-1)$.
Clearly, these two leading divergencies cancel each other.

With this prescription for calculation of $B$-period integrals, one finds
\be
-\oint_{\sqrt{4N}}^\infty \frac{y(z)}{2} dz = -\lim\limits_{\Lambda \rightarrow
\infty} \oint_{\sqrt{4N}}^\Lambda \frac{y(z)}{2} dz =-{\Lambda^2\over 2} -\big( N \log N - N \big) +
2 N \log 2\Lambda +O\left({N\over\Lambda^2}\right)
\ee
\be
\oint_{\sqrt{4N}}^\infty \frac{Ndz}{y^5(z)}  = \oint_0^1
\dfrac{\zeta^3d\zeta}{128N(1-\zeta)^{5/2}}= \dfrac{1}{12N}
\ee
\be
\oint_{\sqrt{4N}}^\infty \frac{21N(z^2+N)}{y^{11}(z)}\, dz = \oint_0^1
\dfrac{21\zeta^7(16-16\zeta+5\zeta^2)}{524288N^3(1-\zeta)^{11/2}}\,d\zeta  = -\dfrac{1}{120N^3}
\ee
{\fontsize{8pt}{0pt}
\be
\oint_{\sqrt{4N}}^\infty \frac{11N(135z^4+558Nz^2+158N^2)}{y^{17}(z)}\, dz =
\oint_0^1 \dfrac{11\zeta^{11}(17280-34560\zeta+30384\zeta^2-13104\zeta^3+2275\zeta^4)}
{1073741824N^5(1-\zeta)^{17/2}}\,d\zeta  = \dfrac{1}{252N^5}
\ee}
so that the first terms of the expansion (\ref{dfdn}) are reproduced.
Clearly, there is an identity
\be
-\oint_B \rho_1(z|N) dz = \dfrac{\partial F_N}{\partial a} + {\rm const}_1
+ {\rm const}_2\cdot N
\label{GeneralGenus}
\ee
i.e. the derivative of the free energy is reproduced modulo $N$-constant and
$N$-linear terms (they can be removed into redefinition/rescaling of
various quantities such as $g$, $\Lambda$ and $Z$ itself).

The simplest way to check this identity in the generic form is to use equation
(\ref{HZeqRecursive}):
\be
\rho_1(z|N+1) - 2 \rho_1(z|N) + \rho_1(z|N-1) =
\dfrac{1}{N} \dfrac{\partial^2}{\partial z^2} \rho_1(z|N)
\ee
This equation directly implies that
\be
\oint_B \rho_1(z|N+1) dz - 2 \oint_B \rho_1(z|N) dz + \oint_B \rho_1(z|N-1) dz =
\dfrac{1}{N} \oint_B \dfrac{\partial^2}{\partial z^2} \rho_1(z|N) dz
\ee
The $B$-contour integral of the full derivative does not vanish, as one could naively
expect. The non-vanishing contribution comes from the genus zero part of the
function:
\be
\dfrac{1}{N} \oint_B \dfrac{\partial^2}{\partial z^2} \rho_1(z|N) =
-\dfrac{1}{N} \oint_B \dfrac{\partial^2}{\partial z^2} \left( \frac{y(z)}{2} \right) dz
= - \dfrac{1}{N}
\ee
All higher genera contributions vanish. Therefore, we proves that
\be
\oint_B \rho_1(z|N+1) dz - 2 \oint_B \rho_1(z|N) dz + \oint_B \rho_1(z|N-1) dz = -\dfrac{1}{N}
\ee
Precisely the same equation is satisfied by the r.h.s. of (\ref{GeneralGenus}):
\be
F_N = \sum\limits_{k = 1}^{N-1} \log k! \ \ \
\Longrightarrow \ \ \ F_{N+1} - 2 F_N + F_{N-1} = \log N! - \log (N-1)! = \log N
\ee
and
\be
\dfrac{\partial}{\partial a} \Big( F_{N+1} - 2 F_N + F_{N-1} \Big) = \dfrac{1}{N}
\ee
Therefore, the l.h.s. and r.h.s. of (\ref{GeneralGenus})
satisfy one and the same equation. This equation has any linear function of
$N$ as a kernel. This completes the proof.

\section*{Appendix III.
Limiting cases of AGT relation}

Conformal blocks and Nekrasov partition functions,
which are identified by the AGT relation,
depend on numerous free parameters
(external and internal dimensions, $\epsilon$'s),
and one can look at various limits at the boundary
of these moduli spaces.
These limits are often highly non-trivial.
When translated into the language of matrix models
they can be also used to interrelate different
$\beta$-ensembles, very much in the spirit of
\cite{AMM.IM}.
In this Appendix we briefly review what is currently
known about these limiting cases, see the following Table.

\bigskip

\hspace{-0.5cm} \centerline{
\begin{tabular}{|c|c|c|c|c|}
\hline
limit & AGT & CFT & SW theory &
$\beta$-ensemble \\
\hline \hline
interior of moduli space & \cite{AGT} & \cite{CFT} & \cite{SW2,Arg}
& DF $\beta$-ensemble \\
\hline $\Delta_{i,ext}\rightarrow\infty$ & \cite{nonconf} &  &
\cite{SW2,Arg,SW1} & DF $\beta$-ensemble if $\kappa\le N_C$\\
$i=1\ldots \kappa$&&&&Unitary $\beta$-ensembles otherwise\\
\hline
$\Delta_{int}\rightarrow\infty$ & \cite{MMMzam,Pogh} & \cite{AlZa}
& perturbative regime & ? \\
\hline
$c \rightarrow \infty$ & \cite{MMc} &  &  & ? \\
($g_s=\sqrt{\epsilon_1\epsilon_2}\rightarrow 0$) &&&&\\
\hline
$\epsilon_2\rightarrow 0$ & \cite{NS,MMbz} & &
quantized SW theory & ? \\
\hline
$\Delta\rightarrow$ degenerate value & \cite{MMu3,MMnf} & \cite{CFT,FL} &  & \cite{Wilmm}\\
\hline
\end{tabular}
}

\bigskip

In the very left column of the Table only the quantities that becomes zero or infinite are
written. For instance, the record $\Delta_{ext}\to\infty$ in the second row
implies that all the external dimensions and the central charge remain finite, while
the record $c\to\infty$ in the fourth row means that all the conformal
dimensions are kept finite.

Originally the AGT conjecture relates conformally invariant
theories in $2d$ and in $4d$.\footnote{Restriction to
$4d$ is actually inessential: dimension can be also
3, 5 or 6.}
This means that one deals with the ${\cal N}=2$ SYM theory
with additional adjoint, bifundamental or fundamental
hypermultiplets, adjusted to guarantee vanishing
of the $\beta$-function (e.g. $N_A=1$ or $N_F=2N_C$
for the simple $SU(N_C)$ theories).

The conformal invariance is broken through the
dimensional transmutation when masses of some
hypermultiplets are taken to infinity.

\bigskip

Another aspect of the AGT conjecture is that it occurs
for the $\epsilon$-deformed SYM theories,
where the $4d$ Lorentz invariance is violated by
peculiar graviphoton backgrounds.
The limit of $g_s = \sqrt{-\epsilon_1\epsilon_2}
\rightarrow 0$, leading to the ordinary SW
theory, is singular.
Remarkably, it is not always a naive planar limit
in the matrix-model language:
of interest are rather peculiar double-scaling limits,
when the DF $\beta$-ensemble reduces to another,
still non-trivial $\beta$-ensemble, not just to
a quasiclassical approximation.

In particular, in the pure gauge limit
what occurs are unitary $\beta$-ensembles \cite{PGL},
which are very interesting and deserve further
investigation.

A special attention is recently attracted to the
double-scaling limit $\epsilon_2\rightarrow 0$
\cite{NS}.
It is interesting, because in this limit
a non-trivial dependence on the single $\epsilon$-parameter,
$\epsilon_1$, survives, despite $g_s\rightarrow 0$,
and the SW theory is quantized in the most naive way:
by switching from the classical integrable system \cite{intSWd}
to its direct quantum counterpart,
and from the spectral curves to the associated Baxter equation.
The SW representation (\ref{SWeq}) for the free
energy now involves {\it exact} (quantum) Bohr-Sommerfeld
periods \cite{MMbz}: the monodromies of the wave
functions of (the Fourier transform of) the Baxter equation, which are simultaneously
non-perturbatively corrected Harish-Chandra functions
\cite{MMMsurop}.
Moreover, the wave function itself has a wonderful
interpretation in terms of CFT:
it is equal to the conformal block with
additional insertion of the degenerate primary \cite{Brav,FLit1,Wylsurf,MT,MMMsurop},
associated with a surface operator insertion \cite{AGTguk} in the
M5/SYM language.
Remarkably, such insertions into the conformal block
can be also made for $\epsilon_2\neq 0$, which corresponds to switching on
Whitham times, \cite{MMMsurop},
still, their exact role in the AGT relation has to be further clarified.
Also an adequate matrix model description of the
Baxter wave function is an open problem.

\end{document}